\documentclass[12pt]{article}

\catcode`\@=11

\global\arraycolsep=2pt
\usepackage{amsbsy,amssymb,latexsym,amsfonts,amsmath}
\usepackage{graphicx,color}
\usepackage{cite}

\begin{document}
\begin{flushright}
\parbox{4.2cm}
{RUP-21-5}
\end{flushright}

\vspace*{0.7cm}

\begin{center}
{ \Large CP-violating supersymmetry anomaly}
\vspace*{1.5cm}\\
{Koichiro Nakagawa and Yu Nakayama}
\end{center}
\vspace*{1.0cm}
\begin{center}

Department of Physics, Rikkyo University, Toshima, Tokyo 171-8501, Japan

\vspace{3.8cm}
\end{center}

\begin{abstract}
We show that CP-violating Weyl anomaly induces a supersymmetry anomaly in the formulation of superconformal supergravity as is observed in CP-preserving cases. This supersymmetry anomaly can be removed in the old minimal supergravity by adding suitable local counterterms, and it becomes a consistent theory.
\end{abstract}

\thispagestyle{empty} 

\setcounter{page}{0}

\newpage





\section{Introduction}

An anomaly is an obstruction to gauge a symmetry. Such an obstruction arises not only in internal symmetries but also in space-time symmetries. If a quantum field theory has a Lorentz anomaly, one cannot consistently couple it to gravity. If a supersymmetric field theory has a supersymmetry anomaly, one cannot consistently couple it to supergravity. Since our paradigm of superstring theory as a consistent theory of quantum gravity relies on (spontaneously broken) supersymmetry, it is disastrous if the supersymmetry is anomalous.

Recently, there has been some renewed interest in an anomaly in supersymmetry. In \cite{Papadimitriou:2019gel}, it was found that when the R-symmetry is anomalous, as is the case in most supersymmetric field theories, the supersymmetry must be simultaneously anomalous by solving the Wess-Zumino consistency condition in superconformal supergravity. The similar observation was made in the context of supersymmetric localization \cite{Genolini:2016sxe,Closset:2019ucb} or AdS/CFT correspondence \cite{Papadimitriou:2017kzw}. Correlation functions and (anomalous) Ward-Takahashi identities have been revisited in perturbation theories \cite{An:2017ihs,Katsianis:2019hhg,Eleftheriou:2020ray,Bzowski:2020tue,Katsianis:2020hzd}. This indicates that the superconformal supergravity (or new minimal supergravity for the same reason \cite{Papadimitriou:2019yug}) is anomalous and inconsistent. Fortunately, we can find a local counterterm to remove the supersymmetry anomaly in the old minimal supergravity, and supergravity revives if we use an appropriate formulation \cite{Katsianis:2020hzd,Bzowski:2020tue} (see also \cite{Kuzenko:2019vvi}).

All these findings are based on the super Weyl anomaly without CP violating terms. Some time ago, it was suggested that in addition to the Euler density and the Weyl tensor squared, the CP-violating Pontryagin density can appear in the Weyl anomaly \cite{Nakayama:2012gu}. There have been some discussions on the explicit demonstration of the Pontryagin density in the Weyl anomaly in 
\cite{Bonora:2014qla,Bonora:2015nqa,Bonora:2015odi,Bonora:2016ida,Bastianelli:2016nuf,Bonora:2018obr,Bastianelli:2018osv,Frob:2019dgf,Bastianelli:2019fot,Bonora:2019dyv,Bastianelli:2019zrq,Abdallah:2021eii}. In our recent paper \cite{Nakagawa:2020gqc}, we have found that the supersymmetric extension of the CP-violating Weyl anomaly can be consistently formulated. This generates new terms in the R-symmetry anomaly, so the structure of the supersymmetry anomaly addressed in the previous paragraph should be reconsidered.

The objective of this paper is to study properties of the supersymmetry anomaly induced by the CP-violating super Weyl anomaly. We solve the Wess-Zumino consistency condition to determine the form of the supersymmetry anomaly. We will further show that this supersymmetry anomaly can be removed in the old minimal supergravity by adding suitable local counterterms, and it becomes a consistent theory.

The organization of this paper is as follows. In section 2, as a warm-up, we study the supersymmetry anomaly induced by a CP-violating flavor anomaly. In section 3, we study the supersymmetry anomaly induced by the CP-violating supersymmetric Weyl anomaly. In section 4, we conclude with some discussions.

\section{CP-violating flavor anomaly}
In this section, we first study a possible supersymmetry anomaly when a global flavor symmetry is anomalous in the context of global supersymmetry. It will turn out that the supersymmetry anomaly is present within the Wess-Zumino gauge, but it can be removed once we are allowed to use the entire supersymmetry multiplet as a local counterterm beyond the Wess-Zumino gauge. Our results agree with \cite{Papadimitriou:2019yug,Bzowski:2020tue}, but we will include more general terms than those considered in the literature.

In this section, we work in the four-dimensional Minkwoski space-time.  Let $\Phi$ be a chiral superfield with the component expansion
\begin{align}
\Phi = \phi + \sqrt{2} \theta \psi + \theta^2 f + \cdots \ .
\end{align}
From the structure of the superspace, the supersymmetry variation of each component is given by\footnote{Throughout the paper, we mostly follow the convention of \cite{Wess:1992cp}.}
\begin{align}
\delta_\epsilon \phi &= \sqrt{2} \epsilon \psi \cr
\delta_{\epsilon} \psi &= i \sqrt{2} \sigma^m \bar{\epsilon} \partial_m \phi + \sqrt{2} \epsilon f \cr
\delta_{\epsilon}f &= i \sqrt{2} \bar{\epsilon} \bar{\sigma}^m \partial_m \psi \ .  \label{rigidv}
\end{align}

In supersymmetric field theories, a $U(1)$ flavor anomaly is summarized by the Konishi anomaly equation \cite{Konishi:1983hf} in superspace
\begin{align}
\bar{D}^2 J = k W^\alpha W_\alpha \ . 
\end{align}
Here, the field strength superfield $W_{\alpha}$ for the background gauge field is a chiral superfield with the component
\begin{align}
W_{\alpha} = -i \lambda_{\alpha} + D\theta_{\alpha} -i F_{mn}(\sigma^{mn})^{\ \beta}_{\alpha} \theta_{\beta} + \theta^2 \sigma_{\alpha \dot{\alpha}}^m \partial_m \bar{\lambda}^{\dot{\alpha}} + \cdots  \ , \label{Wa}
\end{align}
where $F_{mn} = \partial_m A_n - \partial_n A_m$ is the field strength for the flavor gauge field $A_m$ that couples to the anomalous symmetry current $J_m$, and $\lambda$ is the gaugino, and $D$ is an auxiliary scalar.

In the component expansion, the Konishi anomaly equation contains the flavor anomaly.  When $k$ is real, $\partial^m J_m = ik \int d^2\theta W^\alpha W_\alpha  + c.c.$ gives
\begin{align}
\partial^m J_m = \mathrm{Re}(k) (F_{mn} \tilde{F}^{mn} - 2\bar{\lambda}\bar{\sigma}_m   \partial^m \lambda + 2\lambda \sigma^m \partial_m \bar{\lambda} ) \label{1}
\end{align}
Here $\tilde{F}_{kl} = \frac{1}{2}\epsilon_{kl mn} F^{mn}$ as usual.
Note that a gaugino term has been added compared with the standard chiral anomaly. The gaugino term, however, is a total derivative, so it can be attributed to a scheme choice.
If $k$ had an imaginary part, we could include a CP-violating flavor anomaly 
\begin{align}
\partial^m J_m = \mathrm{Im}(k) \left(F_{mn} F^{mn} +2i\bar{\lambda} \bar{\sigma}_m  \partial^m \lambda +  2i\lambda \sigma^m \partial_m \bar{\lambda} - 2 D^2 \right) \  \label{2}
\end{align}
as well. This term is not usually discussed in the literature, but as mentioned in the introduction, such terms may be present when CP is violated \cite{Nakayama:2018dig}.
The supersymmetry anomaly induced by \eqref{1} was discussed in \cite{Papadimitriou:2019yug,Bzowski:2020tue}. We will focus on \eqref{2}.

In the Wess-Zumino gauge, the supersymmetry variation of the background flavor gauge multiplet is
\begin{align}
\delta_\epsilon A_m &= i\epsilon \sigma_m \bar{\lambda} + i \bar{\epsilon} \bar{\sigma}_m \lambda \cr
\delta_\epsilon \lambda & =\sigma^{mn} \epsilon F_{mn} + i \epsilon D  \cr
\delta_\epsilon D & = -\epsilon \sigma^m \partial_m \bar{\lambda} + \bar{\epsilon} \bar{\sigma}^m \partial_m \lambda \ .
\end{align}
The flavor symmetry variation for each is given by 
\begin{align}
\delta_\vartheta A_m &= \partial_m \vartheta  \cr
\delta_\vartheta \lambda &= \delta_\vartheta D = 0 \ . 
\end{align}
We see that these symmetries commute with each other: $[\delta_{\epsilon},\delta_{\vartheta}] =0$.

To see the apparent existence of the supersymmetry anomaly $\mathcal{A}_{\epsilon} = \delta_\epsilon \Gamma$, where $\Gamma$ is the effective action, let us study the Wess-Zumino consistency condition between the supersymmetry variation $\delta_{\epsilon}$ and the flavor symmetry variation $\delta_{\vartheta}$. Under the presence of anomalies, it is given by 
\begin{align}
 \delta_{{\epsilon}} \mathcal{A}_\vartheta -  \delta_\vartheta \mathcal{A}_{{\epsilon}} = [\delta_\epsilon, \delta_{\vartheta}] \Gamma  = 0 \ . 
\end{align}
We will shortly see that the supersymmetry variation of the flavor anomaly $\delta_{{\epsilon}} \mathcal{A}_\vartheta$ is non-zero, so the supersymmetry anomaly $\mathcal{A}_{{\epsilon}}$ cannot vanish (within the Wess-Zumino gauge).

In order to solve the Wess-Zumino consistency condition, it is instructive to realize that the supersymmetry variation of the $\theta^2$ component of a chiral superfield (i.e. $W^\alpha W_{\alpha}$ here) is a total derivative of the  $\theta$ component from \eqref{rigidv}. Since the  holomorphic part of the flavor anomaly $\mathcal{A}_\vartheta$ is given by the $\theta^2$ component of $W^\alpha W_{\alpha}$, the above observation leads to 
\begin{align}
 \delta_{{\epsilon}} \mathcal{A}_\vartheta  =    -2k\int d^4x  \partial_m \vartheta  \bar{\epsilon} \bar{\sigma}^m   (\lambda D -iF_{rs} \sigma^{rs} \lambda ) + c.c.  \ ,
\end{align}
where terms in parenthesis are essentially the $\theta$ component of $W^\alpha W_\alpha$.

This expression is different from that in \cite{Papadimitriou:2019yug,Bzowski:2020tue} because our $\mathcal{A}_\vartheta$ has  extra gaugino terms, but our choice is much more convenient. 
It indicates that the supersymmetry anomaly should contain the form
\begin{align}
\mathcal{A}_\epsilon =  -2k\int d^4x A_m  \bar{\epsilon} \bar{\sigma}^m  (\lambda D -iF_{rs} \sigma^{rs} \lambda )  + c.c.
\end{align}
up to flavor invariant terms. This is again different from what was obtained in  \cite{Papadimitriou:2019yug,Bzowski:2020tue}, but it also satisfies the global supersymmetry Wess-Zumino consistency condition from $[\delta_{\epsilon}, \delta_{\epsilon'}] = 0$.

The appearance of the supersymmetry anomaly may look perplexing, but it is simply an artefact in the Wess-Zumino gauge. It means that once the flavor symmetry is anomalous, one cannot stay in the Wess-Zumino gauge \cite{Kuzenko:2019vvi}. Indeed if we are allowed to use the full vector superfield, one may find that there exists a local counterterm to remove the supersymemtry anomaly. It is easy to find that it is  given by 
\begin{align}
S^{\text{SUSY}}_c = -2i k \int d^4x \chi (\lambda D -iF_{rs} \sigma^{rs} \lambda) + c.c. , 
\end{align}
where $\chi$ is set to zero in the Wess-Zumino gauge but it transforms as $\delta_{\epsilon} \chi = -i \bar{\epsilon} \sigma^m A_m$ under the supersymmetry transformation otherwise. 
Note that the expression for the counterterm is simpler than that in \cite{Bzowski:2020tue}. At this point, we declare that we can choose a scheme in which the supersymemtry is not anomalous under the influence of the $U(1)$ flavor anomaly (if we go beyond the Wess-Zumino gauge).

In our discussions so far, $k$ may be a complex number and the imaginary part gives the CP-violating flavor anomaly. Our conclusion was not affected by the presence of the CP-violating terms.  
When $k$ is real, the difference from our scheme and that in  \cite{Bzowski:2020tue} can be attributed to the local counterterm:
\begin{align}
S_c = k \int d^4x A_m (2{\lambda} \sigma^m \bar{\lambda})  \ . 
\end{align}
The flavor variation of this counterterm explains the extra gaugino terms that we have added in the flavor anomaly. The supersymmetry variation also explains the difference of our supersymmetry anomaly (as well as supersymmetry recovering local counterterms) and those found in the literature \cite{Papadimitriou:2019yug,Bzowski:2020tue}.

\section{CP-violating supersymmetric Weyl anomaly}
Now let us move on to the supersymmetry anomaly induced by a (CP-violating) supersymmetric Weyl anomaly. With all the technical details aside, the presentation will be in parallel with what we have studied in the flavor anomaly in the previous section. From now on, we switch to the superfield formulation of the $\mathcal{N}=1$ supergravity in four dimensions, and the supergravity convention in this section is in accordance with \cite{Wess:1992cp}. We will see that the supersymmetry anomaly is present in the superconformal supergravity, but it can be removed in the old minimal supergravity by an appropriate countereterm.

Let $\Phi$ be a chiral density with the component expansion given by  
\begin{align}
\label{trans}
\Phi = a+ \sqrt{2}\Theta \rho + \Theta^2 f + \cdots \ . 
\end{align}
Under the local supersymmetry transformation, each component of the chiral density $\Phi$ transforms as\footnote{Since we follow the convention of \cite{Wess:1992cp}, we should note that the local supersymmetry variation has an extra minus sign compared with the global one \eqref{rigidv}.}
\begin{align}
\delta_\epsilon a & = -\sqrt{2} \epsilon \rho + \text{gravitino} \cr
\delta_\epsilon \rho &=-\sqrt{2}\epsilon f + (a \ \text{terms})+\text{gravitino} \cr
\delta_\epsilon f &= \partial_m (i\sqrt{2}\rho \sigma^m  \bar{\epsilon} + \text{gravitino}) \ . \label{cdentisy}
\end{align}
In this paper, we will neglect higher order terms in gravitino, so only the relevant terms for our discussions are explicitly presented here. 

The supersymmetric Weyl anomaly is summarized by the superfield equation
\begin{align}
\bar{D}^{\dot{\alpha}} T_{\alpha \dot{\alpha}} + \frac{2}{3} D_{\alpha} T = 0
\end{align}
with a so-called trace chiral superfield $T$. In four dimensions, the Weyl anomaly is given by $a$ coefficient for the Euler density, $c$ coefficient for the Weyl tensor squared as well as $e$ coefficient for the Pontryagin density.\footnote{Our normalizations of curvature invariants are: $\text{Euler} = R^{mnrs} R_{mnrs} - 4 R^{mn} R_{mn} +R^2$, $\text{Weyl}^2 = R^{mnrs} R_{mnrs} - 2R^{mn}R_{mn}+ \frac{1}{3}R^2$, $\text{Pontryagin} = \frac{1}{2} \epsilon_{mnab}R^{mnrs} R^{ab}_{\ \ rs}$.} In the super symmetric field theories, $c$ and $e$ are combined to make a complexified $c$ \cite{Nakagawa:2020gqc}. Since the Euler density term is not affected by the CP violation, our focus will be complexified $c$, where the imaginary part is given by Weyl anomaly coefficient for the Pontryagin density.

In terms of the trace chiral superfield, the supersymmetric Weyl anomaly from the complexified $c$ can be written as
\begin{align}
T = \frac{c}{{16}\pi^2} W^{\alpha\beta\gamma} W_{\alpha\beta \gamma} \ .
\end{align}
The $\Theta^2$ terms of $T$ are given by  the trace of the energy-momentum tensor $T^{m}_{m}$ and the divergence of the R-current $D^m J_m^R$:
\begin{align}
\label{anomaly}
e T^{m}_m &= \frac{c}{32\pi^2}\int d^2\Theta \mathcal{E} W^{\alpha\beta\gamma} W_{\alpha\beta\gamma} + h.c \cr
e D^m J^R_m &=  -\frac{ic}{48\pi^2}\int d^2\Theta  \mathcal{E} W^{\alpha\beta\gamma} W_{\alpha\beta\gamma} + h.c 
\end{align}
Here $W^{\alpha\beta\gamma}$ is the Weyl tensor chiral superfield and $\mathcal{E} = e + \cdots$ is the vielbein superfield. Compared with the conventional supersymmetric Weyl anomaly \cite{Bonora:1985cq,Bonora:2013rta}, we allow the possibility that the coefficient $c$ is a complex number. As we have alluded, this enables us to discuss the case when the Pontryagin density is included in the Weyl anomaly. 

Before we discuss the supersymmetry anomaly, let us briefly review the component form of the supersymmetric Weyl anomaly. The Weyl tensor chiral superfield has the component form 
\begin{align}
W^{\alpha\beta\gamma} = w^{\alpha\beta\gamma} + \Theta^\delta r_{\delta} ^{\ \alpha \beta \gamma} + \cdots \ ,
\end{align}
where  the gravitino field strength is given by $w^{\alpha\beta\gamma} = \sigma^{mn \alpha\beta} D_m \psi_n^\gamma$ in terms of gravitino $\psi_m^\alpha$ (with symmetrizatin on $\alpha \beta \gamma$). The superpartner $r_{\delta} ^{\ \alpha \beta \gamma}$ is given by  $r_{\delta} ^{\ \alpha \beta \gamma} = -3iF^{mn} \sigma_{mn}^{\ \ \ \alpha \beta} \delta^\gamma_{\delta} + \frac{1}{4} R^{mnop} \sigma_{mn}^{\ \ \ \alpha \beta}\sigma_{op \ \delta}^{\ \ \gamma}$ (with symmetrization on $\alpha \beta \gamma$).\footnote{Relevant formulae can be found in  \cite{Lu:2011mw}.}  Here, $F_{mn} = \partial_m A_n -\partial_n A_m$ is the field strength for the R-symmetry, and the Riemann tensor $R_{mnop}$  may be replaced with the Weyl tensor (since the trace part drops out).

To evaluate \eqref{anomaly}, we need to know the $\Theta^2$ component of $W^{\alpha\beta\gamma} W_{\alpha\beta\gamma}$ . Up to higher order terms in gravitino, $W^{\alpha\beta\gamma} W_{\alpha\beta\gamma}$ can be expanded as follows (see e.g. \cite{Lu:2011mw}):
\begin{align}
W^{\alpha\beta \gamma} W_{\alpha\beta \gamma} &=  w^{\alpha\beta\gamma} w_{\alpha\beta \gamma} - 2\Theta^\delta w^{\alpha\beta\gamma} r_{\delta \alpha \beta \gamma}  + \frac{1}{2} \Theta^2 r^{\delta \alpha\beta\gamma} r_{\delta \alpha\beta\gamma}+ \cdots \   \cr
&=
\frac{1}{12}(D_m \psi_n - D_n \psi_m + i\epsilon _{mnop}D^o\psi^p)(D^m \psi^n - D^n \psi^m) \cr
&-2 \Theta^{\delta}\sigma^{mn \alpha\beta} D_m \psi_n^\gamma (-3iF^{op} \sigma_{{op} \alpha \beta} \epsilon_{\gamma \delta} + \frac{1}{4} R^{opqr} \sigma_{{op}  \alpha \beta}\sigma_{{qr} \gamma \delta})   \cr
&+\Theta^2(\text{Weyl}^2-\frac{8}{3}F_{mn} F^{mn}+i\mathrm{Pontryagin}-\frac{8}{3}iF_{mn} \tilde{F}^{mn} )+ \cdots \ .
\end{align}
Note that R-symmetry gauge field $G_m$ in \cite{Lu:2011mw} and $A_m$ in this paper is related by $G_m=\frac{1}{3}A_m$.\footnote{Our convention is R-charge of the supersymmetry generator is one. In the literature, they may use different conventions where R-charge of a conformal fermion is one or that of a conformal scalar is one.} Here, we used the spinor formula $f_{\alpha \beta}f^{\beta}_{\gamma} = \frac{1}{2}\epsilon_{\alpha \gamma} f^{\delta \sigma} f_{\delta \sigma}$.

By substituting this component expression into the superfield anomaly equation (\ref{anomaly}), we obtain
\begin{align}
T^{m}_m =& \frac{\mathrm{Re}(c)}{16\pi^2} \mathrm{Weyl}^2 - \frac{\mathrm{Re}(c)}{6\pi^2}F^{mn}F_{mn} 
 - \frac{\mathrm{Im}(c)}{16\pi^2} \mathrm{Pontryagin} +  \frac{\mathrm{Im}(c)}{6\pi^2}F^{mn}\tilde{F}_{mn} + \text{gravitino}  \cr
D^m J^R_m &=   \frac{\mathrm{Re}(c)}{24\pi^2} \mathrm{Pontryagin} - \frac{\mathrm{Re}(c)}{9\pi^2} F^{mn}\tilde{F}_{mn} + \frac{\mathrm{Im}(c)}{24\pi^2} \mathrm{Weyl}^2 - \frac{\mathrm{Im}(c)}{9\pi^2} F^{mn}F_{mn}  + \text{gravitino} \  \label{superWeylano}
\end{align}
They are in agreement with \cite{Nakagawa:2020gqc}. When $c$ is real, it reproduces the results in \cite{Anselmi:1997am} (with corrections made in \cite{Cassani:2013dba}). Compared with \cite{Papadimitriou:2019gel,Bzowski:2020tue} (when $c$ is real), we have added  the gravitino contribution to the R-symmetry anomaly (as well as to the Weyl anomaly), which will be crucial in the following although the explicit form will not be necessary. Since our focus is CP-violating terms, we have set $a=0$, but inclusion of $a$ with additional gravitino contributions in $D^m J^R_m$ would modify the analysis in \cite{Papadimitriou:2019gel,Bzowski:2020tue}, and possibly facilitate the computation as we will see when $a=0$.

Now, it is time for us to investigate the supersymmetry anomaly induced by the supersymmetric Weyl anomaly \eqref{superWeylano} in the superconformal supergravity. Our strategy is to study the Wess-Zumino consistency condition $\delta_{\epsilon} \mathcal{A}_{\vartheta}- \delta_\vartheta \mathcal{A}_{\epsilon}=0$ between the local supersymmetry $\delta_{\epsilon}$ and the local R-symmetry $\delta_{\vartheta}$, which must satisfy $[\delta_{\epsilon}, \delta_{\vartheta}] = 0$. We then argue that a supersymmetry anomaly $\mathcal{A}_{\epsilon}$ must be present because the supersymmetry variation of the R-symmetry anomaly $\mathcal{A}_{\vartheta}$ is nonzero.

The supersymmetry variation of the R-symmetry anomaly
\begin{align}
\mathcal{A}_{\vartheta} = \int d^4x e \vartheta D^m J_m^R 
\end{align}
 is most easily computed by noting that the  holomorphic part of $\mathcal{A}_{\vartheta}$ is given by the $\Theta^2$ component of the chiral density $\mathcal{E} W^{\alpha\beta\gamma} W_{\alpha\beta\gamma}$. The supersymmetry variation, then, directly follows from \eqref{cdentisy} as a total derivative of the $\Theta$ component of $\mathcal{E} W^{\alpha\beta\gamma} W_{\alpha\beta\gamma}$. With the integration by part, we obtain
\begin{align}
\delta_{{\epsilon}} \mathcal{A}_{\vartheta} =  -\frac{c}{48\pi^2}\int d^4x e (\partial_m \vartheta)  \bar{\epsilon}_{\dot{\delta}} (\bar{\sigma}^m)^{\dot{\delta}\delta} 2w^{\alpha\beta\gamma} r_{\delta \alpha \beta \gamma}  +c.c.\  
\end{align}
up to higher order terms in gravitino. 

The Wess-Zumino consistency condition $\delta_{\epsilon} \mathcal{A}_{\vartheta}- \delta_\vartheta \mathcal{A}_{\epsilon}=0$  therefore predicts the presence of the supersymmetry anomaly $\mathcal{A}_{\epsilon}$. To satisfy the Wess-Zumino consistency condition, it must include

\begin{align}
\mathcal{A}_{{\epsilon}} = -\frac{c}{48\pi^2}\int d^4x e  \bar{\epsilon}_{\dot{\delta}}  A_m (\bar{\sigma}^m)^{\dot{\delta} \delta} 2w^{\alpha\beta\gamma} r_{\delta \alpha \beta \gamma}  +c.c. \label{ssanomaly}
\end{align}
up to possible R-symmetry invariant terms, where the R-symmetry gauge transformation of $A_m$ is $\delta_\vartheta A_m=\partial_m \vartheta$. 

To determine the entire structure of the supersymmetry anomaly, it is instructive to see if the other Wess-Zumino consistency condition is satisfied by \eqref{ssanomaly} . Here, we focus on the non-trivial one from the commutator $[\delta_{\epsilon}, \delta_{{\epsilon'}}] \Gamma = \delta_\vartheta \Gamma$  studied in \cite{Papadimitriou:2019gel},\footnote{We have assumed that there is no diffeomorphism and Lorentz anomaly.} where the gauge parameter is given by $\vartheta =  -2(i\bar{\epsilon}' A_m \bar{\sigma}^m \epsilon - i\bar{\epsilon} A_m \bar{\sigma}^m \epsilon')$. 

The computation of $\delta_{\epsilon'} \mathcal{A}_\epsilon$ is facilitated by realizing that we need to compute the supersymmetry variation of the 
$\Theta$ component of a chiral density (i.e. $w^{\alpha\beta \gamma } r_{\delta \alpha \beta \gamma})$. This is immediately read off from \eqref{cdentisy}. In particular, we may neglect all the gravitino terms as well as terms given by $a$ because $a = w^{\alpha\beta\gamma} w_{\alpha\beta\gamma}$ is a gravitino bilinear.  We then recognize that it is given by the $\Theta^2$ component of the chiral density $\mathcal{E} W^{\alpha\beta\gamma} W_{\alpha\beta\gamma}$. More explicitly, we have
\begin{align}
\delta_{\epsilon'} \mathcal{A}_{\epsilon} = -\frac{2c}{48\pi^2}\int d^4x e  {\bar{\epsilon}  A_m \bar{\sigma}^m \epsilon '} \left(\text{Weyl}^2-\frac{8}{3}F_{rs} F^{rs} +i\mathrm{Pontryagin}-\frac{8}{3}iF_{rs}\tilde{F}^{rs} \right)+c.c. \ 
\end{align}
up to higher order terms in gravitino. Accordingly, the commutator acting on the effective action $\Gamma$ is computed as 
\begin{align}
 [\delta_{\epsilon}, \delta_{{\epsilon}'}] \Gamma &= \delta_{\epsilon} \mathcal{A}_{\epsilon'} - \delta_{\epsilon'} \mathcal{A}_{\epsilon} \cr
& = 
\int d^4x e ( -2(i\bar{\epsilon}' A_m \bar{\sigma}^m \epsilon - i\bar{\epsilon} A_m \bar{\sigma}^m \epsilon') )  \cr
&  \ \ \ \  \ \ \ \left(\frac{\mathrm{Re}(c)}{24\pi^2} \mathrm{Pontryagin} - \frac{\mathrm{Re}(c)}{9\pi^2} F^{rs}\tilde{F}_{rs} + \frac{\mathrm{Im}(c)}{24\pi^2} \mathrm{Weyl}^2 - \frac{\mathrm{Im}(c)}{9\pi^2} F^{rs}F_{rs}  \right) \ 
\end{align}
up to higher order terms in gravitino. This is precisely what we have in the right hand side of the Wess-Zumino consistency condition $ \delta_\vartheta \Gamma = \mathcal{A}_{\vartheta}$ including the CP-violating terms (see \eqref{superWeylano}) with the expected $\vartheta$, so we may regard \eqref{ssanomaly} as a minimal solution of the Wess-Zumino consistency conditions.

Within the superconformal supergravity, there is no local counterterm that can cancel $\mathcal{A}_\epsilon$ and the supersymmetry is hence anomalous. This may sound perplexing, but it simply means that when supersymmetric Weyl transformation is anomalous, one cannot stay in the superconformal supergravity. A conventional way to couple supergravity to supersymmetric field theories without superconformal symmetry is to use the old minimal supergravity by introducing a chiral compensator superfield.

Indeed, in the old minimal supergravity, one may find a suitable local counterterm to cancel the supersymmetry anomaly: 
\begin{align}
S^{\text{SUSY}}_c = -{\frac{c}{32\sqrt{2}\pi^2}}\int d^4x e  \chi^\delta 2w^{\alpha\beta\gamma} r_{\delta \alpha \beta \gamma}   + c.c. \ .
\end{align}
Here $\chi$ is the $\Theta$ component of the chiral compensator of the old minimal supergravity with the supersymmetry variation $\delta_{\epsilon} \chi = -\frac{2\sqrt{2}}{3} A^m \bar{\epsilon} \bar{\sigma}_m$, which may be identified with the spin half component of the gravitino. It is easy to see that this counterterm cancels $\mathcal{A}_\epsilon$ (up to higher order terms in gravitino) and the supersymmetry anomaly can be removed.

Before we conclude, let us mention the difference between our approach and those taken in \cite{Papadimitriou:2019gel}  \cite{Bzowski:2020tue} when $c$ is real. The main difference lies where we have included extra gravitino terms in $\mathcal{A}_\vartheta$ to make the supersymmetry transformation more transparent.
When $c$ is real, this results in the difference of the R-symmetry anomaly given by a total derivative of gravitino bilinears.

The fact that the  difference is a total derivative:
\begin{align}
D^m(J_m^{\text{ours}} - J_m^{\text{theirs}} )=  c D^m G_m \ ,
\end{align}
where $G_m$ is the bilinear of gravitino of the structure $ D\bar{\psi} D\psi$, implies that the gravitino contribution to the R-symmetry anomaly can be generated by a scheme change when $c$ is real. 
 More explicitly it can be attributed to a local counterterm
\begin{align}
 S_c = c \int d^4x e A^m G_m \ . \label{gravc}
\end{align}Then, the difference of our supersymmetry anomaly $\mathcal{A}_\epsilon$ from that in \cite{Papadimitriou:2019gel,Bzowski:2020tue} can be understood as a supersymmetry variation of the same counterterm $S_c$.

We should emphasize that when $c$ is not real, the gravitino contribution to the R-symmetry anomaly is no longer a total derivative, so one cannot eliminate it by using any local counterterm such as in \eqref{gravc}. In this sense, our scheme is more appropriate in the more general cases with CP violations.

\section{Conclusion}

In this paper, we have studied a structure of the supersymmetry anomaly when the supersymmetric Weyl anomaly includes the CP-violating terms proposed in \cite{Nakagawa:2020gqc}. Our presentation is slightly different from those in the literature (even when there is no CP-violating terms) and in our opinion the derivation is technically more transparent. The advantage has come from the addition of higher order terms in gravitino in the R-symmetry anomaly to regard it as a component of a chiral density, which makes the supersymmetry variation of the R-symmetry anomaly easily obtained from the structure of the superfield. It has enabled us to solve the Wess-Zumino consistency condition with no demanding supergravity manipulations in components. 

Within the superconformal supergravity, our results imply that one cannot remove the supersymmetry anomaly (with or without the CP-violating terms). Either one cannot in the new minimal supergravity. This fact is not that surprising because after all when the R-symmetry is anomalous, gauging the R-symmetry which is necessary in these formulations does not make sense (regardless of the existence of the additional supersymmetry anomaly). We have found a suitable local counterterm to remove the supersymmetry anomaly in the old minimal supergravity, where there is no need to gauge the R-symmetry, so coupling to supergravity can be made consistent in this formulation. 

In the main part of this paper, we have not discussed the superconformal anomaly $\mathcal{A}_\eta = \delta_\eta \Gamma$ (i.e. so-called $S$ transformation rather than $Q$ transformation). When $c$ is real, the Wess-Zumino consistency condition with the superconformal transformation has been thoroughly discussed in \cite{Papadimitriou:2019gel}. When $c$ is real, the difference of our $\mathcal{A}_\eta$ and the one in \cite{Papadimitriou:2019gel} should be computed by studying the superconformal variation of \eqref{gravc}. This effectively removes the R-symmetry non-invariant terms of $\mathcal{A}_\eta$ in \cite{Papadimitriou:2019gel}. As a consequence, our superconformal anomaly,  which is essentially the $\Theta$ component of $\mathcal{E}W^{\alpha\beta\gamma} W_{\alpha\beta\gamma}$ (see e.g. \cite{Komargodski:2010rb}), is R-invariant $\delta_\vartheta \mathcal{A}^{\text{our}}_\eta = 0$.  This is in agreement with the Wess-Zumino consistency condition $\delta_\eta \mathcal{A}_\vartheta - \delta_\vartheta \mathcal{A}_\eta=0$ because we have $\delta_\eta \mathcal{A}^{\text{our}}_\vartheta = 0$. The last equality comes from the fact that $\mathcal{A}^{\text{our}}_\vartheta$ is the $\Theta^2$ component of $\mathcal{E} W^{\alpha\beta\gamma} W_{\alpha\beta\gamma}$ which must be superconformal invariant.\footnote{This can be inferred from the fact that the $\Theta^2$ component of $\mathcal{E} W^{\alpha\beta\gamma} W_{\alpha\beta\gamma}$ yields the classical action for the superconformal supergravity.} The consistency of the superconformal anomaly in  more general cases when $c$ is not real will be left for future research.

\section*{Acknowledgements}

This work is in part supported by JSPS KAKENHI Grant Number 17K14301. YN thanks  I.~Papadimitriou and A.~Bzowski for useful correspondence.

\end{document}